\documentclass[doublecol]{epl2} 
\usepackage{geometry}

\usepackage{graphicx,amsfonts,bm,epsfig,color,latexsym}
\usepackage{amsmath}
\usepackage{amsfonts}
\usepackage{amssymb}
\usepackage{textcomp}
\usepackage{lipsum}
\usepackage{epstopdf}
\usepackage[utf8]{inputenc}

\title{Quantum information and statistical complexity of hydrogen-like ions in Dunkl-Schr\"odinger system}
\shorttitle{Title} 

\author{Akash Halder \inst{1} \and Amlan K. Roy\inst{1} \and Debraj Nath\inst{2}}
\shortauthor{F. Author \etal}

\institute{                    
  \inst{1} Department of Chemical Sciences, Indian Institute of Science Education and Research (IISER) Kolkata, Mohanpur-741246, Nadia, WB, India.\\
  \inst{2} Department of Mathematics, Vivekananda College, 269 D.H. Road, Kolkata-700063, WB, India.
}

\abstract{In this work, we present analytical solutions of Schr\"odinger equation for Coulomb potential in presence of a Dunkl reflection operator. Expressions are offered for eigenvalues, eigenfunctions and radial densities for H-isoelectronic series ($Z=1-3$). The degeneracy in energy in absence and presence of the reflection has been discussed. The standard deviation, Shannon entropy, R\'enyi entropy in position space have been derived for arbitrary quantum states. Then several important complexity measures like L\'opez-Ruiz-Mancini-Calbet (LMC), Shape-R\'enyi complexity (SRC), Generalized R\'enyi complexity (GRC), R\'enyi complexity ratio (RCR) are considered in the analytical framework. Representative results are given for three one-electron atomic ions in tabular and graphical format. Changes in these measures with respect to parity and Dunkl parameter have been given in detail. Most of these results are offered here for the first time.}

\begin{document}

\newcommand{\ba}{\begin{array}}
	\newcommand{\ea}{\end{array}}
\newcommand{\bc}{\begin{center}}
	\newcommand{\ec}{\end{center}}
\newcommand{\bds}{\begin{displaymath}}
	\newcommand{\eds}{\end{displaymath}}
\newcommand{\bd}{\begin{document}}
	\newcommand{\ed}{\end{document}}
\newcommand{\ben}{\begin{enumerate}}
	\newcommand{\een}{\end{enumerate}}
\newcommand{\beqa}{\begin{eqnarray}}
	\newcommand{\eeqa}{\end{eqnarray}}
\newcommand{\beqas}{\begin{eqnarray*}}
	\newcommand{\eeqas}{\end{eqnarray*}}
\newcommand{\beq}{\begin{equation}}
	\newcommand{\eeq}{\end{equation}}
\newcommand{\bfg}{\begin{figure}}
	\newcommand{\efg}{\end{figure}}
\newcommand{\bfr}{\begin{flushright}}
	\newcommand{\efr}{\end{flushright}}
\newcommand{\bit}{\begin{itemize}}
	\newcommand{\eit}{\end{itemize}}
\newcommand{\btb}{\begin{tabbing}}
	\newcommand{\etb}{\end{tabbing}}

\newcommand{\lt}{\left}
\newcommand{\rt}{\right}
\newcommand{\lan}{\langle}
\newcommand{\ran}{\rangle}
\newcommand{\la}{\leftarrow}
\newcommand{\ra}{\rightarrow}
\newcommand{\La}{\Leftarrow}
\newcommand{\Ra}{\Rightarrow}
\newcommand{\lefq}{\lefteqn}
\newcommand{\lgla}{\longleftarrow}
\newcommand{\lgra}{\longrightarrow}
\newcommand{\Lgla}{\Longleftarrow}
\newcommand{\Lgra}{\Longrightarrow}
\newcommand{\Lra}{\Leftrightarrow}

\renewcommand{\a}{\alpha}
\renewcommand{\b}{\beta}
\renewcommand{\d}{\delta}
\newcommand{\D}{\Delta}
\newcommand{\ds}{\displaystyle}
\newcommand{\eps}{\epsilon}
\newcommand{\f}{\frac}
\newcommand{\fb}{\framebox}
\newcommand{\g}{\gamma}
\newcommand{\G}{\Gamma}
\newcommand{\hs}{\hspace}
\newcommand{\iny}{\infty}
\newcommand{\lam}{\lambda}
\newcommand{\lb}{\linebreak}
\newcommand{\m}{\mu}
\newcommand{\mc}{\multicolumn}
\newcommand{\mb}{\makebox}
\newcommand{\n}{\nu}
\newcommand{\nab}{\nabla}
\newcommand{\np}{\newpage}
\newcommand{\nn}{\nonumber}
\newcommand{\om}{\omega}
\newcommand{\Om}{\Omega}
\newcommand{\para}{\parallel}
\newcommand{\pad}{\partial}
\newcommand{\p}{\pi}
\newcommand{\pb}{\pagebreak}
\newcommand{\pr}{\prime}
\newcommand{\R}{I\!\!R}
\newcommand{\s}{\sigma}
\newcommand{\Sg}{\Sigma}
\newcommand{\tri}{\triangle}
\newcommand{\Th}{\Theta}
\newcommand{\vare}{\varepsilon}
\newcommand{\vart}{\vartheta}
\newcommand{\vs}{\vspace}
\newcommand{\ze}{\zeta}

\maketitle

\section{Introduction}
The Wigner-Dunkl formulation of quantum mechanics holds relevance in deformation study through quantum algebra and groups \cite{wigner50,yang51}. The quantum observables can be generated from equations of motion by defining a Dunkl operator as a combination of differential-difference and reflection operator as: $\ds D_x f(x) = \frac{d}{d x} f(x) + \frac{\mu}{x} (1-\widehat{R}) f(x)$, where $\mu$ refers to the Wigner parameter. The traditional momentum operator is modified as: $\widehat{p}_x= \ds\frac{\hbar}{i} D_x= \ds\frac{\hbar}{i} \left[ \frac{d}{dx} + \frac{\mu}{x} (1-\widehat{R}) \right]$, while the position-momentum commutation relation is modified as $[\widehat{x}, \widehat{p}_x]=i\hbar(1+ 2\mu \widehat{R})$. When parameter $\mu$ vanishes, this novel algebra leads to Boson algebra. It is connected to two-particle Calogero model if the involved parameter acts as Calogero coupling constant, where relations like $\widehat{R} \widehat{x} = -\widehat{x}\widehat{R}, \ \widehat{R} \widehat{p} = -\widehat{p} \widehat{R}, \ \widehat{R} \widehat{R} =\textbf{I}$ hold, where \textbf{I} refers to the identity operator. In the integrable systems and discrete symmetry quantum models, they play important role because of the interplay of derivative and reflection operator. It can offer newer algebraic structure, spectral properties and introduction of conserved quantities in a quantum system. Moreover, the wave function can offer both continuous and discrete symmetry due to a combination of $\mu$ parameter and $\widehat{R}$ operator. A few variants, such as generalized Dunkl operator, Jacobi Dunkl operator have been discussed in literature. Besides, they also arise in the context of reflection symmetries in multivariable polynomial, integral transform associated with root systems \cite{dunkl2008,dunkl2014}, complete set of quantum integrals \cite{heckman1991}, harmonic analysis, reflection symmetries in finite reflection group (also termed as finite Coexter group), quantum Calogero-Moser-Sutherland model and its generalization \cite{brink1992,lapointe1996}, anyons in (2+1) and (1+1) dimensions \cite{plyushchay1996a}, para-fields and para-statistics \cite{govorkov1983}, fractional statistics, conformal field theory,  various types of oscillators, Coulomb problem, relativistic problem and generalizations such as Dunkl-Laplacian and Dunkl transform. 

The radial and angular components of Dunkl-Schr\"odinger equation (DSE) was exactly solved for free particle, pseudo-harmonic oscillator and Mie-type potentials \cite{mota2022} in spherical coordinate. Analytical solutions \cite{conf2014} were obtained for 3D isotropic quantum harmonic oscillator (QHO) in Cartesian (Hermite polynomial), cylindrical and spherical (radial, angular solutions in terms of Laguerre, Jacobi polynomials) coordinates. The eigensolution of \emph{generalized} DSE for QHO is obtained by taking generalized Dunkl derivative in Cartesian coordinates \cite{dong2023,debraj2024}. The Dunkl-Klein-Gordon (DKG) oscillator wave function can be written in terms of associated Laguerre and Jacobi polynomials---the equations are separable in both Cartesian and spherical coordinates \cite{hamil2022}. The DSE of QHO with time-dependent mass and frequency was considered, in 1D and 3D case, by a Lewis-Riesenfeld method \cite{benchikha2024}. The D-dimensional QHO was analytically solved in Cartesian and polar coordinate \cite{hamil2025}. The superintegrability and dynamical symmetry in Schwinger-Dunkl algebra was analyzed in Dunkl oscillator \cite{conf2014,ghazouani2019}.  

Attempts were made for the DSE in presence of Coulomb potential. The Dunkl-Coulomb potential in a plane is superintegrable and exactly solvable. The constants of motion as well as the symmetry algebra of Hamiltonian was realized through an so(2,1) algebra \cite{genest2015}. The DKG equation for 2D Coulomb potential was studied analytically from a viewpoint of su(1,1) algebra \cite{mota2021}. The Dunkl Coulomb potential in the plane was generalized to a quasi-exactly solvable one (by using a similarity between radial equation and that of the standard Coulomb potential), where sets of (n+1) potentials associated with a given energy are derived \cite{quesne2024}. The radial part of the 2D Dunkl-Coulomb potential in polar coordinates was studied by using su(1,1) Lie algebra and its theory of irreducible representations. The role of parity and Wigner parameter on spatial localization and energy spectrum was presented in detail for 2D DSE within a generalized Dunkl framework \cite{hassanabadi2024}. The analogous 3D problem was also investigated. The coherent states and group theory was utilized to find energy spectrum along with the normalization constant in terms of coherent state parameter \cite{ramirez2018}. Energy spectrum of DKG equation with Coulomb potential was derived analytically \cite{hamil2022}. The DSE in D-dimension was treated analytically, with separation of angular and radial solutions. The former and latter functions were expressed through Jacobi polynomials and confluent hyper-geometric function \cite{hamil2025}. 

Thus while the solution of H-atom problem within Schr\"odinger formulation is practically as old as quantum mechanics, and constitutes a major triumph, the corresponding problem in DSE has been 
published only recently \cite{ghazouani2019}. Moreover it is apparent from the above discussion that, apart from energy spectrum, there is a lack of studies on other properties, including the 
information theoretical measures, of 3D Coulomb potential in a Dunkl framework. Therein lies the primary objective of this communication. Here, we present a detailed investigation on the 
information theoretical analysis of this potential from DSE. At first, Sec.~\ref{sec2.solutions} derives the energy spectrum in terms of eigenfunctions and eigenvalues, for H-like ions, along with 
their densities.  The interesting degeneracy pattern in Dunkl case is 
compared with normal Schr\"odinger treatment of one-electron systems. Then we present expectation values in Sec.~\ref{sec3.linear}, including Shannon entropy and R\'enyi entropy in position space. 
The important complexity measures like L\'opez-Ruiz-Mancini-Calbet (LMC), Shape-R\'enyi complexity (SRC), Generalized R\'enyi complexity (GRC), R\'enyi complexity ratio (RCR) between two density 
distributions are offered in Sec.~\ref{sec4.complexities}. Representative results are given for three one-electron ions, H, He$^+$ and Li$^{2+}$ in tabular and graphical format. Finally, a few 
conclusions are drawn in Sec.~\ref{sec5.con}.   

\section{Exact solution of Dunkl-Schr\"odinger equation for hydrogen-like ions}\label{sec2.solutions}
The time-independent DSE in 3D spherical polar coordinate can be written as,  
\beq\label{se}
\left[-\frac{\hbar^2}{2\mu}\nabla_D^2+V(r)\right]\psi(r,\theta,\phi)=E\psi(r,\theta,\phi),
\eeq 
where $\mu=m_eM_N/(m_e+M_N)$ is the reduced mass, $m_e $  $M_N$ represent electron and nuclear masses respectively, while $\hbar=1.054571817\times10^{-34}\,Js$ is the reduced Planck constant \cite{rmp2025}. The Dunkl-Laplacian operator is expressed as: $\nabla_D^2=M_r-\frac{L_D^2}{\hbar^2r^2}$, where $L_D^2=-\hbar^2\left(N_{\theta}+\frac{1}{\sin^2\theta}B_{\phi}\right)$ is the angular Dunkl momentum operator \cite{conf2014}. Following quantities are defined,
\beq
\ba{ll}
M_r&=\frac{1}{r^{2a}}\frac{\partial}{\partial r}\left(r^{2a}\frac{\partial}{\partial r}\right),~a=1+\mu_x+\mu_y+\mu_z,\\
N_{\theta}&=\frac{1}{\sin\theta}\frac{\partial}{\partial\theta}\left(\sin\theta\frac{\partial}{\partial\theta}\right)-\frac{\mu_z}{\cos^2\theta}(1-\widehat{R}_z)\\
&~~\quad+2\left[(\mu_x+\mu_y)\cot\theta-\mu_z\tan\theta\right]\frac{\partial}{\partial\theta},\\
B_{\phi}&=\frac{\partial^2}{\partial\phi^2}-\frac{\mu_x}{\cos^2\phi}(1-\widehat{R}_x)
-\frac{\mu_y}{\sin^2\phi}(1-\widehat{R}_y)\\
&~~\quad-2\left(\mu_x\tan\phi-\mu_y\cot\phi\right)\frac{\partial}{\partial\phi},
\ea 
\eeq 
where $\mu_x,\mu_y,\mu_z$ are Dunkl parameters. In this article, we have considered a one-electron atom with nucleus of charge $Zq$, where the corresponding potential energy function is:
\beq
V(r) = -\frac{Z q^2}{4\pi\eps_0 r}, 
\eeq
where $q=1.602176634\times10^{-19}\,C$ is the elementary charge, $\eps_0=8.8541878188\times10^{-12}\, FM^{-1}$, the vacuum electric permittivity in SI unit and the atomic number $Z=1, 2, 3, · · ·$ represent H atom, singly ionized He atom, doubly ionized Li atom, and so on.

The Eq.~(\ref{se}) has solution in the form: $\psi(r,\theta,\phi)=\frac{R(r)}{r^a}H(\theta)G(\phi)$, where angular functions $G,H$ are written as \cite{conf2014}:
\beq
\ba{ll}
G_{m}=N_{m}^{(\phi)} \sin^{e_2}\phi \cos^{e_1}\phi P_{m- \frac{2-s_1-s_2}{4}}^{(   \mu_y - \tfrac{s_1}{2},  \mu_x - \tfrac{s_2}{2})}(\cos2\phi),\\
H_{\ell,m} =N_{\ell,m}^{(\theta)} \cos^{2m} \theta \cos^{e_3}\theta P_{\ell- \frac{1-s_3}{4}}^{( 2m + \mu_x + \mu_y,  \mu_z - \frac{s_3}{2})}(\cos2\theta),
\ea 
\eeq
$e_i=(1-s_i)/2$ and the corresponding radial wave function can be obtained as \cite{goldman1961}, 
\beq
R_{n_r,\ell,m}(r) = \ds N^{(r)}_{n_r,\ell,m}\, (\g r)^{L+1} e^{-\g r/2} L_{n_r}^{(2L+1)}(\g r),
\eeq
where $P_{n}^{(\a,\b)} (x)$ is the Jacobi polynomial of degree $n$ with parameters $\a$, $\b$, and $L_{n}^{(\a)}(x)$ is the associated Laguerre polynomial of degree $n$ with parameter $\a$. The corresponding normalization constants are expressed as: 
\beq
\ba{ll}
(N_{m}^{(\phi)})^2 =\frac{(2m + \mu_x + \mu_y) (m - \frac{2 -s_1- s_2}{4})!\Gamma(m + \mu_x + \mu_y + \frac{2-s_1 - s_2}{4})}{\Gamma(m + \mu_x  + \frac{2- s_1+s_2}{4}) \Gamma(m + \mu_y + \frac{2 +s_1- s_2}{4})},\\
(N_{\ell,m}^{(\theta)})^2 = \frac{(2\ell + 2m +a - \tfrac{1}{2})!(\ell + \frac{1 - s_3}{4})!\Gamma(\ell + 2m + a- \frac{1 +s_3}{4})}{\Gamma(\ell + \mu_z + \frac{3 - s_3}{4})\Gamma(\ell + 2m + \mu_x + \mu_y  + \frac{3+ s_3}{4})},\\
(N_{n_r,\ell,m}^{(r)})^2=\frac{n!\g}{2(n+L+1)\G(n+2L+2)},
\ea 
\eeq 
where $\G(x)$ denotes the Gamma function. The separation constants are given by: $\eps^{(\phi)}_{m}= -4m \left( m + \mu_x + \mu_y \right)$, and $\eps^{(\theta)}_{\ell,m}= -\left(2\ell+2m \right) \left(2\ell+2m + 2a-1\right)$. These represent eigenvalues corresponding to angular momentum operators for $\phi$ and $\theta$, where $L(\ell,m)=2\ell+2m+\mu_x+\mu_y+\mu_z$. Note that, in the normal Schr\"odinger system without reflection $L=\ell$, where $\ell, m, n_r$ are termed orbital, magnetic and radial quantum numbers. The constant $\g$ is found as: $\gamma=\ds{2Z}/{[a_0(n_r+L+1)^2]}$, where $a_0=4\pi\eps_0\hbar^2/[\mu q^2]$ is the Bohr radius. The energy can be expressed as {\color{red}\cite{ghazouani2019}:} $E_{n_r,\ell,m}^{(DD)}(\mu_x,\mu_y,\mu_z) = \ds-\mu Z^2\a^2 c^2/[2(n_r+2\ell+2m+a)^2]$, where $n_r=0,1,2,\cdots;~\ell(s_3)=\{0,0.5,1,1.5,\cdots,\ell^*\}; m(s_1,s_2)=\{0,0.5,1,1.5,\cdots,m^*\}$ and $\alpha=\ds q^2/(4\pi\eps_0\hbar c)$ is fine structure constant. It is dimensionless having value equal to $7.2973525643\times10^{-3}$, while $c=299792458\,m/s$ is the speed of light. The subscript ``DD" signifies results obtained by engaging Dunkl derivative or using DSE. It can also be written in the form: $E_{n_r,\ell,m}^{(DD)}=-\ds E_H/[2(n_r+2\ell+2m+a)^2]$ where $E_H=\ds\mu \a^2c^2$ is the Hartree energy. Note that, $\ell=0$ only when $s_3=1$ whereas $m=0$ only when $s_1=s_2=1$. In general, there exists $k_1,k_2\in\mathbb{N}$ such that $\ell(s_3)=k_1-1+(1-s_3)/4$ and $m(s_1,s_2)=k_2-1+(2-s_1-s_2)/4$. In absence of reflections with usual/ordinary derivative (OD) the energy can be written as \cite{goldman1961} $E_{n_r,\ell,m}^{(OD)} = \ds-\mu Z^2\a^2 c^2/[2(n_r+\ell+1)^2]$, $n_r=0,1,2,\cdots;~\ell=0,1,2,\cdots,n-1;~m=-\ell,\cdots,0,\cdots,\ell$, where $n$ is the principal quantum number such that $n=n_r+\ell+1$. In Dunkl system, $L$ is a non-negative real number, in general. If $\mu_x+\mu_y+\mu_z=k$, $k\in\mathbb{N}$, then we can construct $n$ as $n=n_r+L+1$. In the special case of $\mu_x+\mu_y+\mu_z=0$, the Dunkl system has significant effect of reflection in the solutions. In this case it is easy to construct $n,\ell,m$ for Dunkl case and compare with the usual quantum system without reflections. In Table~\ref{table1.ell}, we have shown the maximum possible values of $\ell^*$ and $m^*$ in the Dunkl system, in the middle and right segments. 

\begin{table}[h]
	\caption{\label{table1.ell} Maximum possible values of $\ell^*, m$ in terms of $n$ (in the left) and $m^*, \ell$ in terms of $n$ (in the right) in the Dunkl plane, $\mu_x+\mu_y+\mu_z=0$ with reflections. }
	\begin{center}
	\centering
	\scalebox{.65}{\begin{tabular}{ccc|ccc|ccc}\hline\hline
			\multicolumn{3}{c}{Parity} & & \multicolumn{2}{c}{$n$}  &   &  \multicolumn{2}{c}{$n$}  \\ \hline 
			$s_3$ & $s_2$ & $s_1$ & $m$ & $\ell^*$ (even $n$) &$\ell^*$(odd $n$) & $\ell$ & $m^*$ (even $n$) &  $m^*$ (odd $n$)\\\hline
			$+$ & $+$ & $+$ & 0 & --- & $\ds(n-1)/2$ & 0 &--- &$\ds(n-1)/2$ \\
			$+$ & $+$ & $-$ & 1/2 & $\ds(n-2)/2$ & ---  & 0 & $\ds(n-1)/2$ & --- \\
			$+$ & $-$ & $+$ & 1/2 & $\ds(n-2)/2$ & ---  & 0 & $\ds(n-1)/2$ & --- \\
			$+$ & $-$ &$-$ &1& --- &$\ds(n-3)/2$&0& --- &$\ds(n-1)/2$\\\hline			
			$-$ & $+$ & $+$ & 0 &  $\ds(n-1)/2$ & ---  & 1/2&$\ds(n-2)/2$ & --- \\
			$-$ & $+$ & $-$ & 1/2 &--- &$\ds(n-2)/2$&1/2& ---   &$\ds(n-2)/2$  \\
			$-$ & $-$ & $+$ & 1/2 &--- &$\ds(n-2)/2$&1/2& --- &$\ds(n-2)/2$\\
			$-$ & $-$ & $-$ & 1 &  $\ds(n-3)/2$& --- &1/2&$\ds(n-2)/2$& --- \\ \hline\hline		
	\end{tabular}}
	\centering
	\end{center}
\end{table}

\subsection{Density function and energy degeneracy of H-like ions in Dunkl-Schr\"odinger system}
The probability density of a state $\psi_{n_r,\ell,m}^{(s_1,s_2,s_3)}$ in DS system is given as: $\rho_{n_r,\ell,m}^{(s_1,s_2,s_3)}(r,\theta,\phi)=|\psi_{n_r,\ell,m}^{(s_1,s_2,s_3)}(r,\theta,\phi)|^2$ such that, $\int_{r=0}^{\infty}\int_{\theta=0}^{\pi}\int_{\phi=0}^{2\pi}p_{n_r,\ell,m}^{(s_1,s_2,s_3)}(r,\theta,\phi)d\chi=1$, where $d\chi=d\chi_{\phi}d\chi_{\theta}d\chi_r$, $d\chi_r=r^{2a}dr$, $d\chi_{\theta}=|\sin\theta|^{2\mu_x+2\mu_y}|\cos\theta|^{2\mu_z}\sin\theta\,d\theta$, $d\chi_{\phi}=|\cos\phi|^{2\mu_x}|\sin\phi|^{2\mu_y}d\phi$, are weighted measures along $\phi,\theta, r$ axes. The ground state of an electron in the lowest orbital is given as: $\psi_{0,0,0}^{(1,1,1)}=\ds N_{0,0,0}^{(r)}\g^a\exp\{-Z\xi/a\}/\sqrt{2B(\mu_x+\mu_y+1,\mu_z+1/2)}$ $\times\sqrt{B(\mu_y+1/2,\mu_x+1/2)}$, where $\xi=r/a_0$, with energy $E^{(DD)}_{0,0,0}=-\mu Z^2\a^2 c^2/(2a^2)$. Then if $P^{(DD)}_{0,0,0}(r)dr$ represents the probability of finding an electron between regions $r$ and $r+dr$, one writes $P^{(DD)}_{0,0,0}(r)dr=2B(\mu_x+\mu_y+1,\mu_z+1/2)B(\mu_y+1/2,\mu_x+1/2)|\psi_{0,0,0}^{(1,1,1)}|^2r^{2a}dr$ and, therefore, $P^{(DD)}_{0,0,0}(r)=[N_{0,0,0}^{(r)}]^2  \left(2Z\xi/a\right)^{2a}\exp\{-2Z\xi/a\}$. On the other hand, in absence of reflection, the probability of finding the electron between radius $r$ and $r+dr$ is $P^{(OD)}_{0,0,0}(r)=[N_{0,0,0}^{(r)}]^2 (2Z\xi)^{2}e^{-2Z\xi}$ \cite{goldman1961}. Considering these, the probability distributions of a state with $n_r$ number of nodes is: $P_{n_r,0,0}^{(DD)}(r)=[N_{n_r,0,0}^{(r)}]^2 \left(2Z\xi/a\right)^{2a}\exp\{-2Z\xi/a\} L_{n_r}^{(2a-1)}\left(2Z\xi/a\right)$, and $P_{n_r,0,0}^{(OD)}(r)=[N_{n_r,0,0}^{(r)}]^2 (2Z\xi)^{2}e^{-2Z\xi} L_{n_r}^{(3)}\left(2Z\xi\right)$. One observes that, in the Dunkl plane $\mu_x+\mu_y+\mu_z=0$, $P_{n_r,0,0}^{(DD)}(r)=P_{n_r,0,0}^{(OD)}(r)$, as expected. 
\begin{figure*}[h]
	\centering
	\includegraphics[width=17cm,height=10cm]{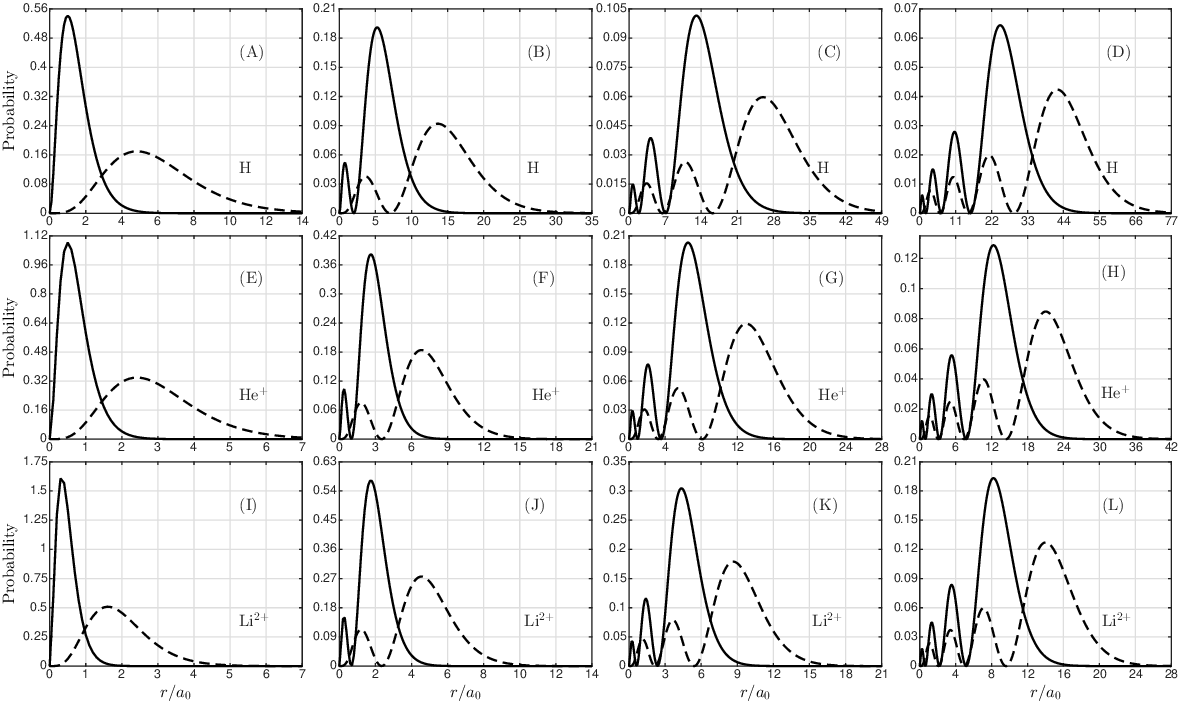}
	\caption{\label{fig1.radial} Probability of finding the electron in between $r$ to $r+dr$. The first, second and third rows, from top, are given for H, He$^{+}$ and Li$^{2+}$. The first, second, third and fourth columns from left are represented for $(n_r,\ell,m)=(0,0,0), (1,0,0), (2,0,0)$ and $(3,0,0)$ states. Solid lines correspond to ``no reflection" ($\mu_x=\mu_y=\mu_z=0$), whereas dashed lines refer to reflections with Dunkl parameters $(\mu_x,\mu_y,\mu_z)=(0.4,0.5,0.3)$.}
\end{figure*}
In Fig.~\ref{fig1.radial} we have plotted the probability distributions of H, He$^+$, Li$^{2+}$ ions for $(\mu_x, \mu_y, \mu_z)=(0.4, 0.5, 0.3)$, in ground and some low-lying states, having quantum numbers $(n_r,\ell,m)=(0,0,0), (1,0,0), (2,0,0)$ and $(3,0,0)$ with and without reflections. From this figure, we observe that with an increase in Dunkl parameters $\mu_x, \mu_y, \mu_z$, the distribution function extends farther in the $x$ axis. While the number of radial nodes remains same in both cases, the peak positions shift to the right side acquiring larger values of $r$. In our recent study \cite{dn.arxiv} on ro-vibrational energy and information analysis of Deng-Fan molecular potential in diatomic molecules, it has been shown that reflection operators have direct impact on angular wave functions $H_{\ell,m}^{(s_3)}$ and $G_m^{(s_1,s_2)}$, and indirect influence on radial wave function $R_{n_r,\ell,m}$. The quantum numbers $m$ and $\ell$ depend on the parity values $s_1,s_2$, $s_3=\pm1$ of reflection operators. The latter operators act in spherical coordinate as, $\widehat{R}_x\psi(r,\theta,\phi)=\psi(r,\theta,\pi-\phi)$, $\widehat{R}_y\psi(r,\theta,\phi)=\psi(r,\theta,-\phi)$, $\widehat{R}_z\psi(r,\theta,\phi)=\psi(r,\pi-\theta,\phi)$, whereas they act in Cartesian coordinate as follows, $\widehat{R}_xf(x,y,z)=f(-x,y,z)$, $\widehat{R}_yf(x,y,z)=f(x,-y,z)$, $\widehat{R}_zf(x,y,z)=f(x,y,-z)$.

\begin{figure*}[h]
	\centering
	\includegraphics[width=5.75cm,height=5cm]{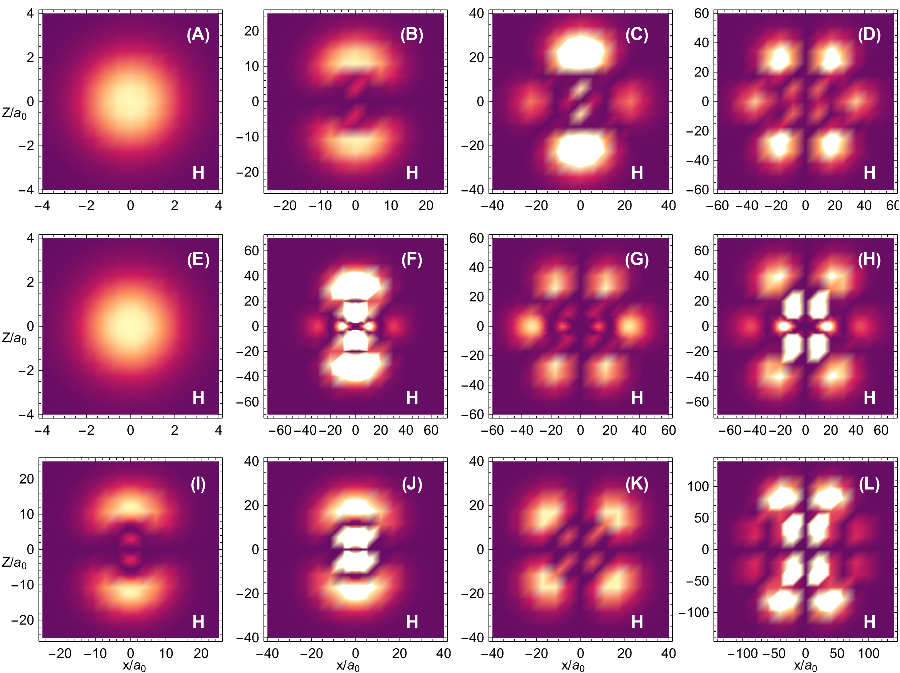}
	\includegraphics[width=5.75cm,height=5cm]{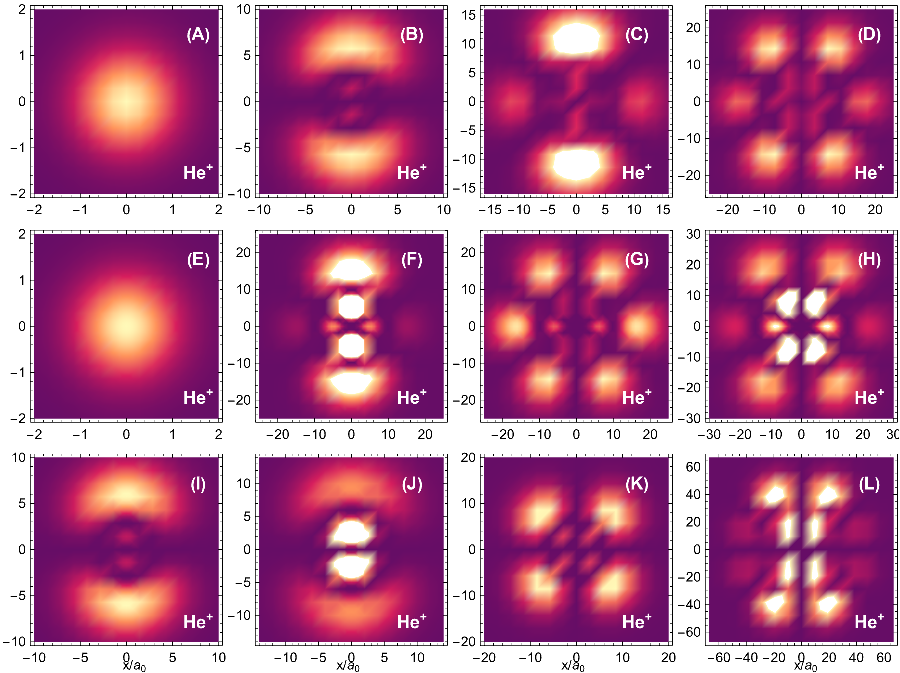}
	\includegraphics[width=5.75cm,height=5cm]{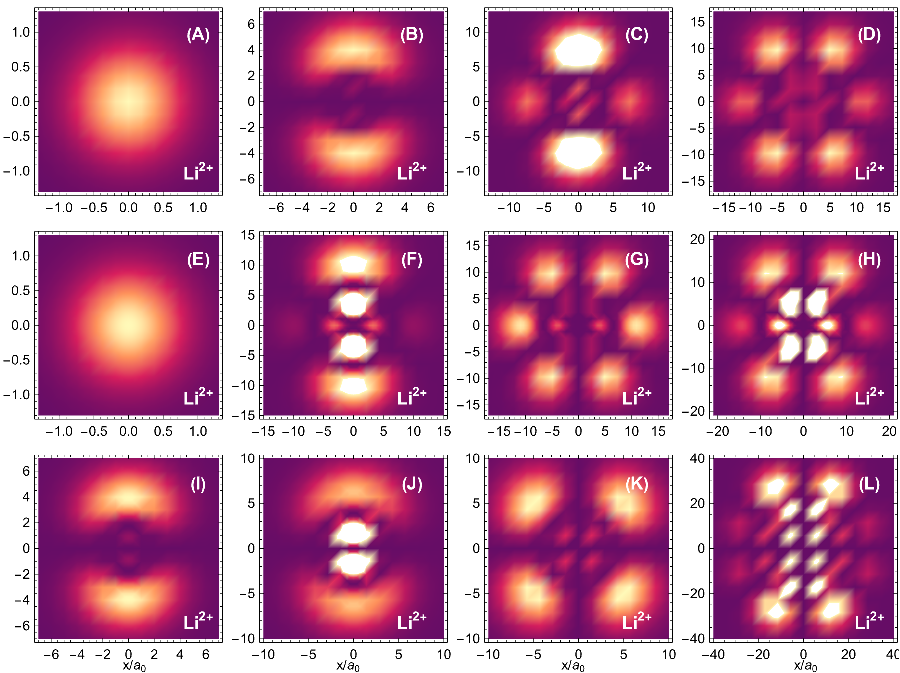}
	\caption{\label{fig2.density} Probability densities of H, He$^+$ and Li$^{2+}$, in $y=1$ plane. The quantum numbers $(n_r, \ell,m)$ are (A) $(0,0,0)$, (B) $(1,1,0)$, (C) $(1,2,0)$, (D) $(1,3,0)$. (E) $(0,0,0)$ (F) $(1,1,1/2)$ (G) $(1,1,1/2)$ (H) $(1,1,1)$ (I) $(1,1/2,0)$ (K) $(1,1/2,1/2)$ (J) $(1,1/2,1/2)$ (L) $(2,3/2,1)$. For (A)-(D), 	$(\mu_x,\mu_y,\mu_z)=(0,0,0)$ and for (E)-(L) the parities are $+++,++-,+-+,+--,-++,-+-,--+,---$ respectively with Dunkl parameters $(\mu_x,\mu_y,\mu_z)=(0.1,-0.2,0.1)$. }
\end{figure*}

\begin{figure*}[h]
	\centering
	\includegraphics[width=5.75cm,height=5cm]{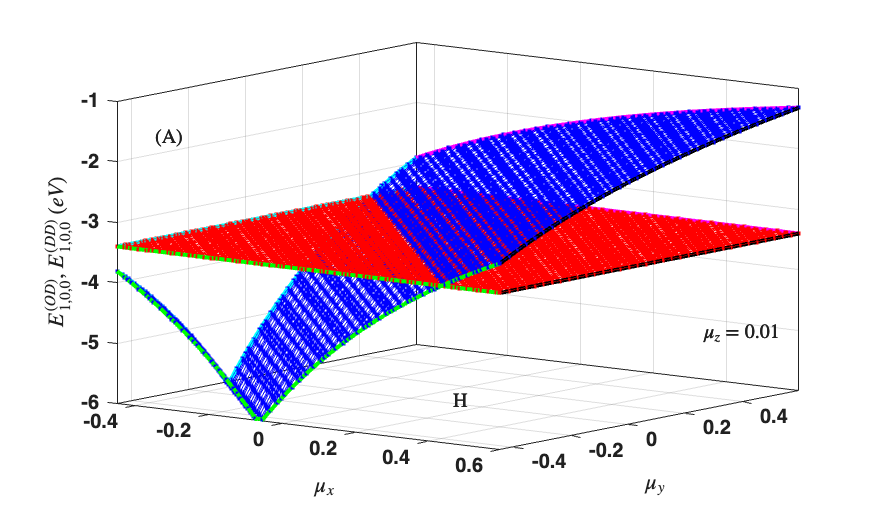}
	\includegraphics[width=5.75cm,height=5cm]{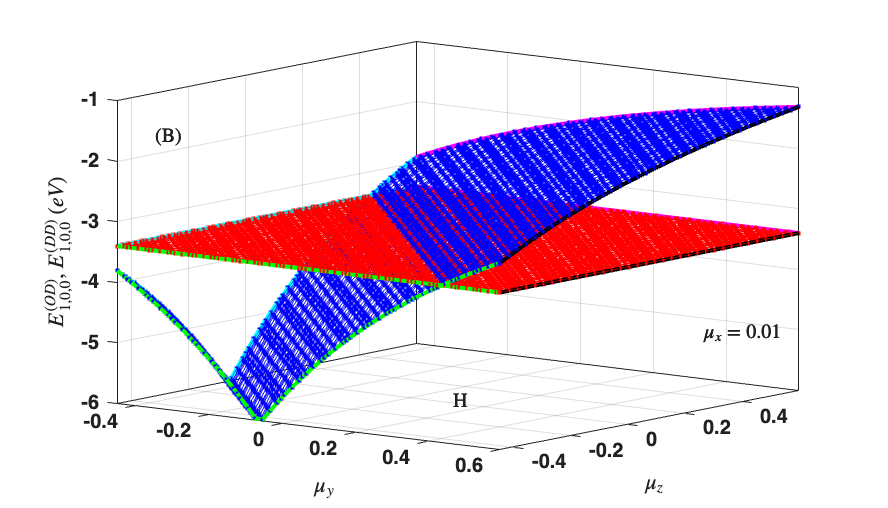}
	\includegraphics[width=5.75cm,height=5cm]{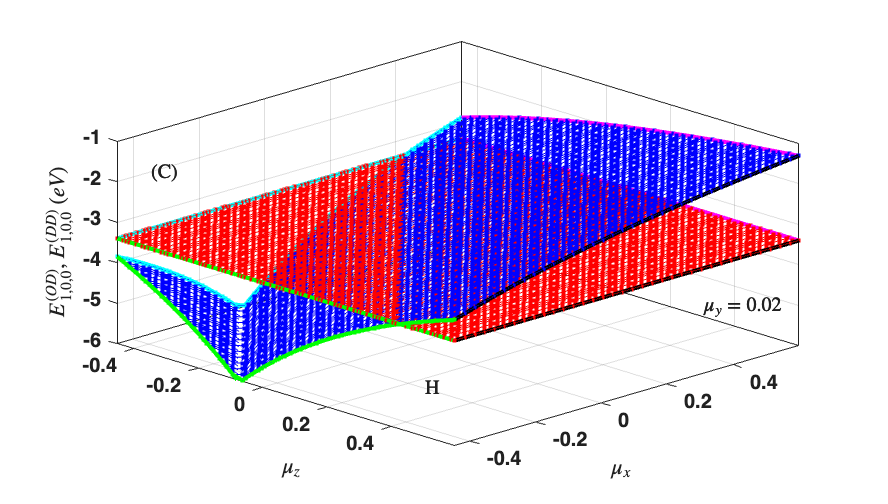}
	\caption{\label{fig3.energy} The degeneracy of first excited state of H atom for $\ell=m=0$, with reflections having parities $+++$ (blue surface) and without reflections (red plane). Panel (A) corresponds to that with respect to $\mu_x$, $\mu_y$; (B) with respect to $\mu_y$, $\mu_z$; (C) with respect to $\mu_z$, $\mu_x$. The red 	plane has same energy value of $E_{1,0,0}^{(OD)}=-3.3995731\,eV$ of H atom, without reflection, in the usual Schr\"odinger system.}
\end{figure*}

To realize the effect of reflections on density functions, the total densities of three ions are plotted in Fig.~\ref{fig2.density}, in the $y=1$ plane. The first four columns from left represent H, the next four for He$^+$ and the last four columns offer results for Li$^{2+}$. For each of these systems, the first row represents density functions without reflection, whereas the second and third rows offer the same in presence of reflection. The respective state quantum numbers, as well as Dunkl parameters are indicated in the caption of figure. The Dunkl plane segment $\mu_x+\mu_y+\mu_z=0$ passes through the origin $(0,0,0)$, and at the origin there are no reflections, and thus usual derivatives are applied. Leaving aside the origin, in the Dunkl region $\mathbb{R}^{(D)}=\{(\mu_x,\mu_y,\mu_z):\mu_x,\mu_y,\mu_z>-\frac{1}{2}\}$ reflection operators have significant role in determining the nature of wave functions; hence Dunkl derivatives are used instead of normal derivatives. 
To find the electronic energy, we have considered three ions with atomic number $Z=1,2,3$. The atomic masses and corresponding reduced masses, taken from \cite{prohaska2022,data} are collected in Table~\ref{table2.parameters}. In presence of reflection, the energy is defined as $E_{n_r,\ell,m}^{(DD)}(\mu_x,\mu_y,\mu_z) = \ds-\mu Z^2\a^2 c^2/[2(n_r+2\ell+2m+a)^2]$, whereas in its absence, it is given by $E_{n_r,\ell,m}^{(OD)} = \ds-\mu Z^2\a^2 c^2/[2(n_r+2\ell+2m+1)^2]$, and independent of quantum numbers $\ell,m$. However, as evident, $E_{n_r,\ell,m}^{(DD)}$ depends on both $\ell,m$. It is known that, for each $\ell$ there are $2\ell+1$ degenerate states having same energy value $E_{n_r,\ell,m}^{(OD)}$. The calculated energies of various states of three species in absence and presence of Dunkl derivative are gathered in Table~\ref{table3.energy}. The numbers in the parentheses signify non-Dunkl energies, obtained from exact expression $\ds \frac{-Z^2}{2n^2}$ a.u. (1 a.u. $=$ 27.2113834 eV). For a fixed set of Dunkl parameters two states, $\psi_{n_r',\ell',m'}^{(s'_1,s'_2,s'_3)}$ and $\psi_{n_r^{''},\ell^{''},m^{''}}^{(s^{''}_1,s^{''}_2,s^{''}_3)}$ have same energy if $n_r'+2\ell'(s'_3 )+2m'(s'_1,s'_2)=n^{''}_r+2\ell^{''}(s''_{3})+2m^{''}(s''_{1},s''_{2})$. The energy $E_{n_r',\ell',m'}^{(DD)}(\mu_x,\mu_y,\mu_z)$ is equal to $E_{n_r^{''},\ell^{''},m^{''}}^{(OD)}$ in the Dunkl plane segment $\mu_x+\mu_y+\mu_z=p$, bounded by the Dunkl region $\mathbb{R}^{(D)}=\{(\mu_x,\mu_y,\mu_z):\mu_x,\mu_y,\mu_z>-\frac{1}{2}\}$ where $p=n_r^{''}+\ell^{''}-(n_r'+2\ell'+2m')$, $\ell'\in\{0,1/2,1,3/2,\cdots,\ell^*\}$, $m'\in\{0,1/2,1,3/2,\cdots,m^*\}$, $\ell^{''}=0,1,2,\cdots$; $n_r',n_r^{''}=0,1,2,\cdots$. The Dunkl plane segment cuts the axes at (p,0,0), (0,p,0) and (0,0,p) and its distance from origin is $p/\sqrt{3}$.
In Fig.~\ref{fig3.energy}, we have plotted the first excited state with quantum numbers $\ell=m=0$ of H atom (in eV) for $+++$ parity state with respect to  $\mu_x$, $\mu_y$ and $\mu_z$. The red planes denote the energy value $E_{1,0,0}^{(OD)}=-3.3995731\,eV$ of H atom, without reflection, in the usual Schr\"odinger system. From this figure, one can observe that there are infinitely many values of Dunkl parameters for which $E_{1,0,0}^{(DD)}=-3.3995731\,eV=E_{1,0,0}^{(OD)}$. Therefore, one can find a suitable set of Dunkl parameters from the Dunkl plane $\mu_x+\mu_y+\mu_z=p,~p\ge0$ for which degenerate energy level may exist.

\section{Linear information measures of Coulomb potential in Dunkl-Schr\"odinger system}\label{sec3.linear}
\subsection{Standard deviation \label{sec3.1rms}}
The expectation value, $\left\langle r^j\right\rangle_{n,\ell,m}^{(DD)}= \int_0^{\infty} r^jR^2_{n_r,\ell,m}(r)dr$ in the DS system can be written analytically as: 
\beq \label{exprj}
\ba{ll}
\left\langle r^j\right\rangle_{n_r,\ell,m}^{(DD)}
= \ds\frac{\Gamma(2L+2+n_r) [N_{n_r,\ell,m}^{(r)}]^2}{n_r! \gamma^{j+1}}  \sum\limits_{i=0}^{2n_r} \frac{2\widetilde{B}_{2+i,2} x_{ij}}{(2+i)!},
\ea
\eeq
$x_{ij}=\Gamma[2L+3+i+j]$ for $j>-3$,  
and $\widetilde{B}_{m,\ell}(c_0^{(n_r,\ell)},\cdots,(i+1)!c_i^{(n_r,\ell)})$ denotes the Bell polynomial where 
\beq
c_i^{(n_r,\ell)}=\left\{\ba{ll}  \ds\frac{(-1)^i\binom{n_r}{i}\sqrt{\Gamma[n_r + 2L(\ell) + 2]}}{\Gamma[2L(\ell)+ i + 2]\sqrt{n_r!}}, &i\le n_r\\0,&i>n_r\ea\right\}.
\eeq The expectation of $r$ of electron in a nodeless state is given by: $\left\langle r\right\rangle_{0,\ell,m}^{(DD)}=\ds a_0(2L+2)(2L+3)/[4Z]$. The standard deviation of $r$ for ground state for arbitrary $\ell,m$ quantum number is: $(\Delta r)_{0,\ell,m}^{(DD)}=\sqrt{\left\langle r^2\right\rangle_{0,\ell,m}-\left\langle r\right\rangle_{0,\ell,m}^2}=\ds a_0(2L+2)\sqrt{2L+3}/(4Z)$ and therefore has the ratio $\left[\left(\Delta r\right)_{0,\ell,m}/\left\langle r\right\rangle_{0,\ell,m}\right]^{(DD)}=1/\sqrt{4\ell(s_3)+4m(s_1,s_2)+2a+1}$, $\left[\left(\Delta r\right)_{0,\ell,m}/\left\langle r\right\rangle_{0,\ell,m}\right]^{(OD)}=1/\sqrt{2\ell+3}$. It is clear that the ratio decreases as $\ell$ increases and the minimum value exists for the highest allowed orbital angular momentum $\ell=\ell^*$ for DD and $\ell=n-1$ for OD. The minimum value is $\left[\left(\Delta r\right)_{0,\ell,m}/\left\langle r\right\rangle_{0,\ell,m}\right]^{(OD)}_*=1/\sqrt{2n+1}$. For DD the minimum value $\left[\left(\Delta r\right)_{0,\ell,m}/\left\langle r\right\rangle_{0,\ell,m}\right]^{(DD)}_*$ depends on $n$, $m(s_1,s_2)$ \& $s_3$ and the corresponding $\ell^*$ are offered in Table~\ref{table1.ell}. There are $n$ number of quantum states having minimum ratio $\left[\left(\Delta r\right)_{0,\ell,m}/\left\langle r\right\rangle_{0,\ell,m}\right]^{(DD)}_*=1/\sqrt{2n+1}$ for the ground state has no nodes in radial wave function such that $n-2(m+\ell)-1=0$.

\begin{figure}[h]
	\centering
	\includegraphics[width=8.5cm,height=5cm]{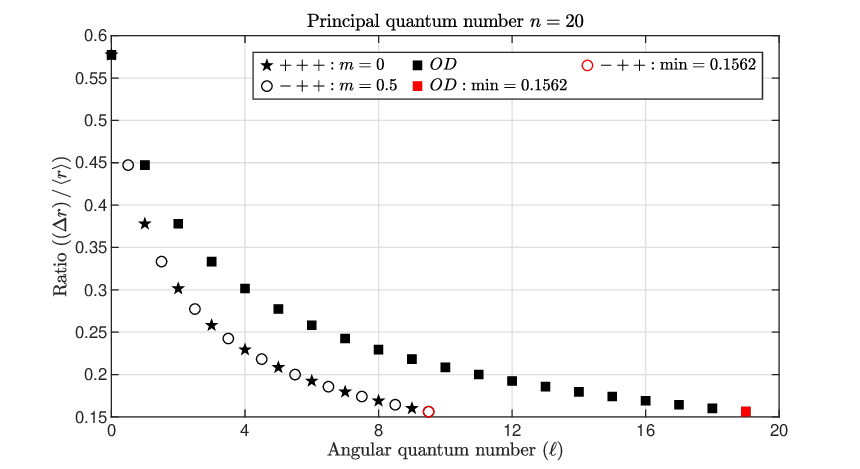}
	\caption{\label{fig4.ratio} Plot of the ratio, $\left[\left(\Delta r\right)_{0,\ell,m}/\left\langle r\right\rangle_{0,\ell,m}\right]^{(DD)}$ of H atom in Dunkl plane $\mu_x+\mu_y+\mu_z=0$, with reflection, having parities $+++$ (star), $-++$ (circle), and without reflection (square) 	$\left[\left(\Delta r\right)_{0,\ell,m}/\left\langle r\right\rangle_{0,\ell,m}\right]^{(OD)}$. The red circle and red square indicate minimum values of corresponding ratios.}
\end{figure}

Figure~\ref{fig4.ratio} plots the ratio $\left[\left(\Delta r\right)_{0,\ell,m}/\left\langle r\right\rangle_{0,\ell,m}\right]^{(DD)}$, in the Dunkl plane $\mu_x+\mu_y+\mu_z=0$ with $m=0$, where reflection operator $\widehat{R}_x,\widehat{R}_y$ have even parity, i.e., $s_1=s_2=1$, while having two different values of (i) $s_3=1$ (star) and (ii) $s_3=-1$ (circle). The same ratio without Dunkl operator (reflection) is given by: $\left[\left(\Delta r\right)_{0,\ell,m}/\left\langle r\right\rangle_{0,\ell,m}\right]^{(OD)}$ (box). One observes that $\left[\left(\Delta r\right)_{0,\ell,m}/\left\langle r\right\rangle_{0,\ell,m}\right]^{(DD)}_*= \left[\left(\Delta r\right)_{0,\ell,m}/\left\langle r\right\rangle_{0,\ell,m}\right]^{(OD)}_*=0.1562$.

\begin{table}[h]
		\centering\caption{\label{table2.parameters} Set of parameters \cite{prohaska2022,data} in SI unit \cite{rmp2025}.  $^{\dagger}$$1\, amu=1.66053906660\times10^{-27}\,Kg$., 	$m_{e}=5.48579909065\times10^{-4}\,amu$. $x=\times10^{-27},y=\times10^{-31}$}
	\begin{center}
		\begin{tabular}{llrr}\hline\hline
		 Atom & Atomic mass  & Atomic mass &Reduced mass \\
		& $M_N(amu)$ & $M_N(Kg)$&$\mu(Kg)^{\dagger}$\\ \hline			
			H  & 1.008 & $1.67382337x$ & $9.10443y$\\
			He & 4.0026& $6.64647699x$& $9.10814y$\\
			Li & 6.94 &$11.52414112x$&$9.10866y$\\ \hline\hline
	\end{tabular}
	\end{center}
\end{table}
\subsection{Shannon entropy\label{sec3.2shannon}}
Now, we discuss Shannon entropy in $\xi\left(\mathbb{R}^*\times[0,\pi]\times[0,2\pi],d\chi\right)$ space, in a given state $\psi_{n,\ell,m}^{(s_1,s_2,s_3)}$, which is given by, 
\beq
\ba{ll}
\mathcal{S}_{n_r,\ell,m}^{(s_1,s_2,s_3)}=-\ds\int\rho_{n_r,\ell,m}^{(s_1,s_2,s_3)}(\mathbf{r})\ln[\rho_{n_r,\ell,m}^{(s_1,s_2,s_3)}(\mathbf{r})]d\chi.
\ea 
\eeq 
It can be written as $\mathcal{S}_{n_r,\ell,m}^{(s_1,s_2,s_3)}=\mathcal{S}_{n_r,\ell,m}^{(r)}+\mathcal{S}_{\ell,m}^{(\theta,s_3)}+\mathcal{S}_m^{(\phi,s_1,s_2)}$, where $\mathcal{S}_{n_r,\ell,m}^{(r)}$, $\mathcal{S}_{\ell,m}^{(\theta,s_3)}$ and $\mathcal{S}_{m}^{(\phi,s_1,s_2)}$ are Shannon entropies of marginal density functions (functions of $r, \theta, \phi$). The quasi-analytical form of radial entropy,  $\mathcal{S}_{n_r,\ell,m}^{(r)}=-\int_0^{\infty} \rho_{n_r,\ell,m}^{(r)}\ln\left[\rho_{n_r,\ell,m}^{(r)}\right]d\chi_r$ is derived as: 
\beq\label{shannon.r}
\mathcal{S}_{n_r,\ell,m}^{(r)}= \frac{N_{n_r,\ell,m}^2}{\g}
[\sum\limits_{i=0}^{2n_r} \frac{2\widetilde{B}_{2+i,2}f(i)}{(2+i)!}-{S}^{(1)}]- \ln[ N_{n_r,\ell,m}^{2} \g^{2a}],
\eeq
${S}^{(1)}=\int_{0}^{\infty} e^{-x}x^{2L+3}[ L_{n_r}^{(2L+1)}(x) ]^{2}\ln[ L_{n_r}^{(2L+1)}(x)]^{2}dx$. The 

\noindent analytical value of ${S}^{(1)}$ is obtained in ref.~\cite{dn.arxiv} by factorization method but quasi-analytical expressions are reported in ref. \cite{yanez1994} and numerical results provided  in ref. \cite{roy2018a}. Thus the value of $\mathcal{S}_{n_r,\ell,m}^{(r)}$ can be obtained from Eq.~(\ref{shannon.r}). The term $f(i)$ is given as, $f(i)=\Gamma(2L+3+i)\left[(2a-2L-2) \Psi_d(2L+3+i)\right. \left.+\G(2L+3+i)\right]$, where $\Psi_d(x)$ refers to the digamma function. Similarly, the Shannon entropy $\mathcal{S}^{(\theta,s_3)}_{\ell,m}$ of angular density function $\rho_{\ell,m}^{(\theta,s_3)}(\theta)$ is found as follows: $\mathcal{S}^{(\theta,s_3)}_{\ell,m}=-\int_{0}^{\pi}\rho_{\ell,m}^{(\theta,s_3)}\ln[\rho_{\ell,m}^{(\theta,s_3)}]d\chi_{\theta}$. Lastly, the Shannon entropy $\mathcal{S}^{(\phi,s_1,s_2)}_m$ of density function $\rho_{m}^{(\phi,s_1,s_2)}$ can be expressed as: $\mathcal{S}^{(\phi,s_1,s_2)}_m=-\int_{0}^{2\pi}\rho_{m}^{(\phi,s_1,s_2)}\ln[\rho_{m}^{(\phi,s_1,s_2)}(\phi)]d\chi_{\phi}$.

The analytical expressions of Shannon entropies of marginal density functions, $\rho_{\ell,m}^{(\theta,s_3)}$ and $\rho_{m}^{(\phi,s_1,s_2)}$ have been produced in \cite{dn.arxiv}. These entropies are independent of $Z$. From Eq.~(\ref{shannon.r}) we observe that the first term $\ln\left[  (N_{n_r,\ell,m}^{(r)})^{2} \g^{2a}\right]$ depends on $Z$. Therefore, for a fixed set of quantum numbers $(n_r,\ell,m)$, Shannon entropy of one-electron ions having $Z>1$ in the DS case, deviates by the amount $[(2a+1)\ln(Z)]_{Z=2,3,\cdots}$, from that of H atom. Without reflection the deviated amount is $[3\ln(Z)]_{Z=2,3,\cdots}$. As we know, the entropy has significant contribution in statistical complexity, and those will be discussed in appropriate places in a future section (Sec.~\ref{sec4.complexities}).

\begin{table}[h]
	\caption{\label{table3.energy} Some low-lying energy levels of H, He$^+$, Li$^{2+}$, in eV, in Dunkl plane $\mu_x+\mu_y+\mu_z=0$ with reflection, having parity values $s_1 ,s_2,s_3=\pm1$.  $1\,J=6.241509074\times10^{18}\,eV$. With usual derivatives the degenerate energy values are same for $^{\dagger} (n_r,\ell,m)$ = (0,0,0);~~ $^{\ddagger} (n_r,\ell,m)$ = (1,0,0), (0,1,0), (0,1,$\pm$1);~~ $^{\S} (n_r,\ell,m)$ = (2,0,0), (1,1,0), (1,1,$\pm$1), (0,2,0), (0,2,$\pm$1), (0,2,$\pm$2);~~ $^{\mathparagraph}(n_r,\ell,m)$ = (3,0,0), (2,1,0), (2,1,$\pm$1), (1,2,0), (1,2,$\pm$1), (1,2,$\pm$2), (0,3,0), (0,3,$\pm$1), (0,3,$\pm$2), (0,3,$\pm$3).}
	\begin{center}
	\scalebox{.8}{
		\begin{tabular}{lcc|c|rrr}\hline\hline
			$n_r$&$\ell$ &$m$&  $s_3s_2s_1$& \multicolumn{1}{c}{H} & \multicolumn{1}{c}{He$^+$}& \multicolumn{1}{c}{Li$^{2+}$} \\ \hline	
				0  &  0 & 0 & $+++$ & 13.59829$^{\dagger}$ & 54.41531$^{\dagger}$ & 122.44156$^{\dagger}$\\\hline
			0 &  0 & 1/2 &  $+\pm\mp$ & 3.39957$^{\ddagger}$ & 13.60383$^{\ddagger}$ & 30.61039$^{\ddagger}$\\	
			0 &  1/2 & 0 & $-++$ & 3.39957 & 13.60383 & 30.61039\\
			1 &  0 & 0 & $+++$ & 3.39957 & 13.60383 & 30.61039\\\hline
			0 &  0 & 1 & $+\pm\pm$ & 1.51092$^{\S}$ & 6.04615$^{\S}$ & 13.60462$^{\S}$\\
			0 &  1/2 & 1/2 & $-\pm\mp$& 1.51092 & 6.04615 &13.60462\\	
			0 &  1 & 0 & $+++$& 1.51092 & 6.04615  &13.60462\\
			1 &  1/2 & 0 & $-++$ &1.51092 & 6.04615  &13.60462\\
			1 &  0 & 1/2 &+ $\pm$ $\mp$&1.51092& 6.04615  &13.60462\\
			2 &  0 & 0 & $+++ $&1.51092 & 6.04615 & 13.60462\\\hline
			0 &  1/2 & 1 & $---$ & 0.84989$^{\mathparagraph}$ & 3.40096$^{\mathparagraph}$ & 7.65260$^{\mathparagraph}$\\\hline\hline
	\end{tabular}}
	\end{center}
\end{table}

The calculated Shannon entropies of H atom in Dunkl system are graphically shown in the top row of Fig.~\ref{fig5.shannon}, for $n=7$ with $\ell=0,1/2,1,3/2,2,5/2,3$ and $m=0,1/2,\cdots,(n-2\ell-1)/2$. For each $n$, there are $n^2$ numbers of Shannon information. For a representative $n$ ($=7$), this figure depicts all possible entropies for eight parities $+++,++-,+-+,+--,-++,-+-,--+,---$, separated by the vertical dashed lines. In each panel, all allowed $m$ quantum numbers are presented in $x$ axis. For $s_1=s_2=1$, $m=0,1,2,3$; for $s_1=s_2=-1$, $m=1,2,3$; and for $s_1=\pm1,s_2=\mp1$, $m=1/2,3/2,5/2$. In each state, the entropy reaches the lowest value at $m=(n-2\ell-1)/2$, which indicates $n_r=0$. Therefore, in each segment of top row of this figure, the lowest value of entropy is obtained for the ground state having $\ell=0$, while it assumes the highest value at $m=0$, implying that $n_r$ is maximum. Therefore, for each $n$, entropy reaches the global maximum if $\ell=0,m=0$. Also, one notices that, it has local minimum value if $m=(n-2\ell-1)/2$. For a fixed state, it increases as $n_r$ increases. 

\subsection{R\'enyi entropy}\label{sec3.3renyi-entropy} 
The entropic moment of density function $\rho_{n_r,\ell,m}^{(s_1,s_2,s_3)}$ is expressed as: $\mathcal{I}_{n_r,\ell,m}^{(s_1,s_2,s_3,\a)}=\int[\rho_{n_r,\ell,m}^{(s_1,s_2,s_3)}(\mathbf{r})]^{\a}d\chi=\mathcal{I}_{n_r,\ell,m}^{(r,\a)}\mathcal{I}_{\ell,m}^{(\theta,\a)}\mathcal{I}_m^{(\phi,\a)}$, $\a>0$, where $\mathcal{I}_{n_r,\ell,m}^{(r,\a)}=\int_0^{\infty}\left[\rho_{n_r,\ell,m}^{(r)}\right]^{\a}d\chi_r$, $\mathcal{I}_{\ell,m}^{(\theta,s_3,\a)}=\int_0^{\pi}\left[\rho_{\ell,m}^{(\theta,s_3)}\right]^{\a}d\chi_{\theta}$ and $\mathcal{I}_m^{(\phi,s_1,s_2,\a)}=\int_0^{\pi}\left[\rho_{m}^{(\phi,s_1,s_2)}\right]^{\a}d\chi_{\phi}$ are the individual entropic moments of marginal density functions. The analytical expressions of radial entropic moments can be obtained for positive integral order $(\a\in\mathbb{N})$ as: 
\beq
\ba{lr}
\mathcal{I}_{n_r,\ell,m}^{(r,\a)}
&=\frac{\Gamma(n_r +2L+2)[(N_{n_r,\ell,m}^{(r)})^{2}]^{\a}}{n_r!\gamma^{2a(1-\a)+1}}\sum\limits_{i=0}^{2n_r\a} \frac{(2\a)! \widetilde{B}_{2n_r\a+i,2\a}}{(i+2\a)! }.\\
&\frac{\Gamma \left[ (2L + 2)\a + 2a(1 - \a) + i + 1 \right]}{\a^{\,\left[(2L + 2)\a + 2a(1 - \a) + i + 1\right]}}
\ea 
\eeq
The analytical expressions for $\mathcal{I}_{\ell,m}^{(\theta,s_3,\a)}$ and $\mathcal{I}_m^{(\phi,s_1,s_2,\a)}$ have been reported in our recent work \cite{dn.arxiv}. Note that $\mathcal{I}_{\ell,m}^{(\theta,s_3,\a)}$ and $\mathcal{I}_m^{(\phi,s_1,s_2,\a)}$ are independent of $Z$, but $\mathcal{I}_{n,\ell,m}^{(r,\a)}$ depends on $Z$. Therefore, for fixed $(n_r,\ell,m)$ values, the entropic moment of H-isoelectronic series $(Z \geq 2)$ is scaled by the amount $[Z^{(2a+1)(1-\a)}]_{Z=2,3,\cdots}$ from the hydrogenic counterpart, in the Dunkl case. Without reflection (non-Dunkl case), the same scaled factor equals $[Z^{3(1-\a)}]_{Z=2,3,\cdots}$.

Now, the R\'enyi entropy of order $\a$, of the density function $\rho_{n,\ell,m}^{(s_1,s_2,s_3)}$ can be expressed as: 
\beq
\mathcal{R}_{n,\ell,m}^{(s_1,s_2,s_3,\a)}=\frac{1}{1-\a}\ln\left[\mathcal{I}_{n,\ell,m}^{(s_1,s_2,s_3,\a)}\right],\a>0,\ne 1.
\eeq 
This can be written as a sum of the corresponding entropies of marginal density functions given as follows: $\mathcal{R}_{n,\ell,m}^{(s_1,s_2,s_3,\a)}=\mathcal{R}_{n,\ell,m}^{(r,\a)}+\mathcal{R}_{\ell,m}^{(\theta,s_3,\a)}+\mathcal{R}_m^{(\phi,s_1,s_2,\a)}$, where $\mathcal{R}_{n,\ell,m}^{(r,\a)}=\ln\left[\mathcal{I}_{n,\ell,m}^{(r,\a)}\right]/(1-\a),$ $\mathcal{R}_{\ell,m}^{(\theta,s_3,\a)}=\ln\left[\mathcal{I}_{\ell,m}^{(\theta,s_3,\a)}\right]/(1-\a)$, $\mathcal{R}_m^{(\phi,s_1,s_2,\a)}=\ln\left[\mathcal{I}_m^{(\phi,s_1,s_2,\a)}\right]/(1-\a)$. 

\begin{figure*}[h]
	\centering
	\includegraphics[width=17cm,height=10cm]{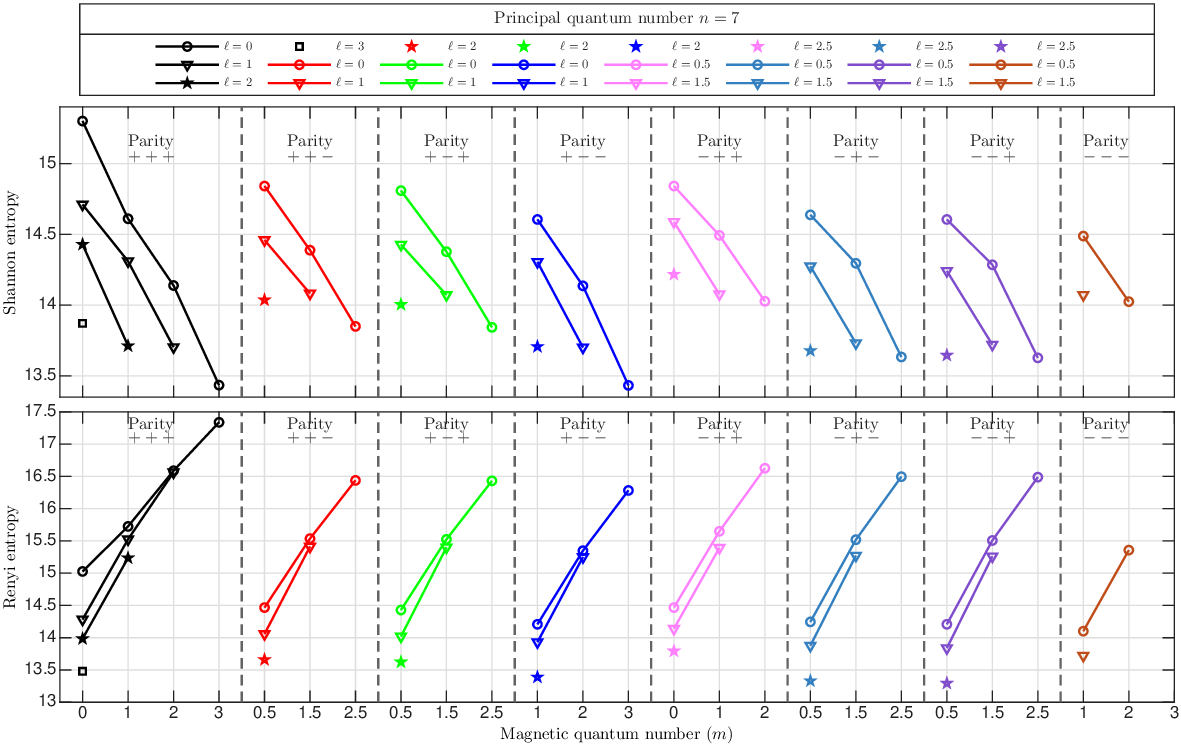}
	\caption{\label{fig5.shannon} Shannon and R\'enyi entropies of H atom with respect to $m$ in DS system. The connected lines join the points with different colors for ease of understanding. All quantities in a.u. The parameters are $(\mu_x,\mu_y,\mu_z)=(0.01,-0.02,0.01)$ and for R\'enyi entropy order $\alpha=1.5$.}
\end{figure*}

The calculated R\'enyi-entropies of H atom are plotted in the bottom row of Fig.~\ref{fig5.shannon}, for $n=7$, having $\ell=0,1/2,1,3/2,2,5/2,3$ and $m=0,1/2,\cdots,(n-2\ell-1)/2$ and order $\a=3/2$. The corresponding quantum numbers considered in this case are the same as that for Shannon entropies. For order $\alpha=3/2$, in each state the entropy reaches highest value at $m=(n-2\ell-1)/2$. From each segment, we observe that maximum entropy is obtained for the ground state. On the other hand, it attains the lowest value for the highest excited state. Therefore, for each $n$, R\'enyi entropy is global minimum if $\ell=0,m=0$ and in each case, it is local maximum, when $m=(n-2\ell-1)/2$. For the order $\alpha>1$ the nature of entropy is opposite to that for $0<\alpha<1$.
\section{Statistical complexity of Coulomb potential in DS system}\label{sec4.complexities}
\begin{figure*}[h]
	\centering
	\includegraphics[width=17cm,height=10cm]{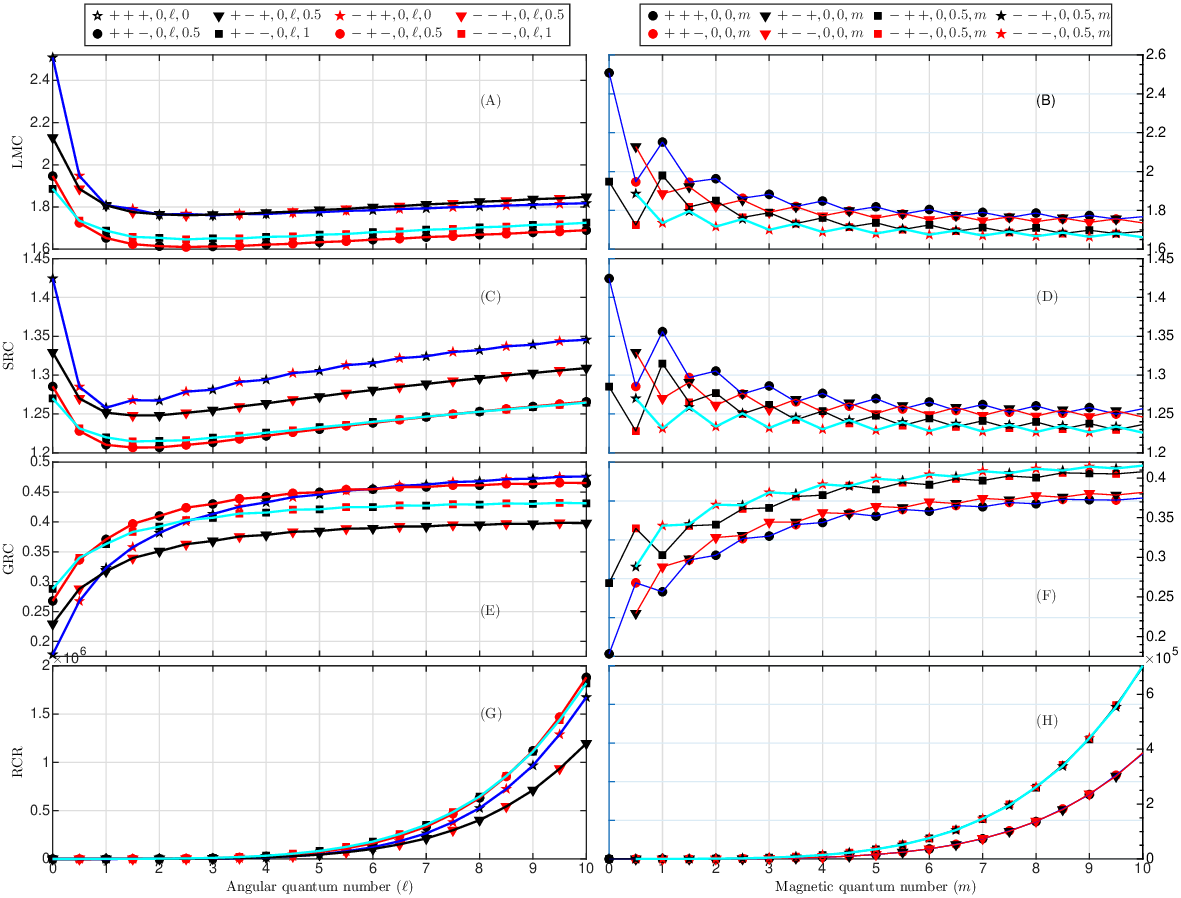}
	\caption{\label{fig6.lmc} Comparison of statistical complexities (in atomic units) for hydrogen atom of the total density function in absence and presence of reflections with 	respect to $\ell$ and $m$. In (A), (B) LMC $C_{0,\ell,m}^{(DD,LMC)}$, (C), (D) SRC $C_{0,\ell,m}^{(DD,SRC,1.5)}$, (E), (F) GRC $C_{0,\ell,m}^{(DD,GRC,1.5,0.5)}$ and (G), 	(H) RCR $C_{(0,\ell,m),(0,0,0)}^{(DD,RCR,1.5,0.5)}$. The parameters are $(\mu_x,\mu_y,\mu_z)=(0.1,-0.2,0.1)$. The connected are drawn for a better appreciation of the figure.}
\end{figure*}

The LMC complexity is defined as: \cite{lmc14,lmc15,lmc16,coulson22,lmc17,lmc.hydrogen.atom,dehesa2009epjd,dn.lmc} $C_{n_r,\ell,m}^{(DD,LMC)}=I_{n_r,\ell,m}^{(DD,2)}$ $\times\exp\left\{\mathcal{S}_{n_r,\ell,m}^{(DD)}\right\}$, where $I_{n_r,\ell,m}^{(DD,2)}$ is disequilibrium or often called second-order entropic moment or self similarity \cite{coulson16}. This can be obtained from previous subsection \ref{sec3.3renyi-entropy} as: $I_{n_r,\ell,m}^{(DD,2)}=e^{-\mathcal{R}_{n_r,\ell,m}^{(DD,2)}}$. Note that, in the limiting case $\a\rightarrow1$, R\'enyi entropy reduces to Shannon entropy, $\mathcal{S}_{n_r,\ell,m}^{(s_1,s_2,s_3)}$. Therefore LMC can be written as: $C_{n_r,\ell,m}^{(DD,LMC)}=\ds\lim\limits_{\a\rightarrow1}\exp\left\{\mathcal{R}_{n_r,\ell,m}^{(DD,\a)}-\mathcal{R}_{n_r,\ell,m}^{(DD,2)}\right\}$. Since R\'enyi entropy is a decreasing function of order $\alpha$, one finds that $C_{n_r,\ell,m}^{(DD,LMC)}>1$, in general, for a bound state. But for highly excited state (having large number of nodes) disequilibrium goes to zero, whereas R\'enyi and Shannon entropies both go to infinity, and so LMC tends to become zero. Note that, LMC is zero for perfect order or maximal disorder \cite{lmc14,montgomery2008}. The numerically calculated values of LMC of H are provided in Table~\ref{table4.complexity} along with the quantum numbers and parities. It is well known that LMC decreases as $\ell$ increases; also for a given $n$, its minimum value occurs at $\ell=n-1$. Note that, without reflection ($\mu_x=\mu_y=\mu_z=0$), the obtained LMC values match with those for the Schr\"odinger case, offered in \cite{lmc.hydrogen.atom}. For all three systems, the magnitude remains same, as it is independent of $Z$. Hence it is provided only for H atom, for a few states, in Fig.~\ref{fig6.lmc} (A), in DS system for $n=2\ell+2m+1$, $\ell=0,1/2,1,3/2,\cdots,10$ and possible minimum value $m=0$ for $(++)$, $m=1/2$ for $(+-)$, as well as $(-+)$, and $m=1$ for $(--)$. Eight parities $(s_3,s_2,s_1)$, namely, $+++, ++-, +-+, +--, -++, -+-, --+, ---$ are considered and the corresponding $n_r,\ell,m$ are shown in the legend. We observe that in DS system, LMC decreases as $\ell$ increases and the respective minimum exists at the possible largest value of $\ell(s_3)=(n-2m(s_1,s_2)-1)/2$. It is also found that, for $-+-$ parity state, LMC records the lowest value of $1.610265$ in case of $\ell=5/2$, $m=1/2$ and $n=7$. Similarly, LMC values are compared with respect to $m(++,+-,-+,--)=0-12;\,1/2-12;\,1/2-12;\,1-12$ for $n=2\ell+2m+1$ with possible minimum value $\ell(\pm)=0(+),1/2(-)$ in panel (B) of Fig.~\ref{fig6.lmc}. We see that, LMC decreases as $m$ increases and the corresponding minimum exists at possible highest value $m(s_1,s_2)=(n-2\ell(s_3)-1)/2$. This is clear from panels (A), (B) of Fig.~\ref{fig6.lmc} that the reflection operators have significant effect in LMC. 

\begin{table}[h]
	\caption{\label{table4.complexity} Statistical complexities ($C_{n_r,\ell,m}^{(DD,LMC)}$, $C_{n_r,\ell,m}^{(DD,SRC,\a)}$, $C_{n_r,\ell,m}^{(DD,GRC,\a,\b)}$ and $C_{(n_r,\ell,m),(0,0,0)}^{(DD,RCR,\a,\b)}$) of H atom, in a.u., for $(\mu_x,\mu_y,\mu_z)=(0.01,-0.015,0.01)$. For SRC $\a=3/2$, and for GRC, RCR $\a=3/2,\b=1/2$.}
	\begin{center}
	\scalebox{.725}{
		\begin{tabular}{lcc|c|c|c|c|c}\hline\hline
			\multicolumn{3}{c|}{Quantum no.}&\multicolumn{1}{c|}{Parity}& LMC &    SRC$^{(\a)}$ &      GRC$\ds{}^{(\a,\b)}$       &    RCR     \\\hline
			$n_r$&$\ell$ &$m$&  $s_3s_2s_1$  & H                          &       H         &       H           &  H                    \\ \hline	
			0  &  0 & 0 & $+++$           & 2.518411 &                1.425505 &       0.176957   &  0.176957         \\ \hline
			0  &  0 & 1/2 & $++-$        & 1.999445 &                1.297449 &       0.256070   &  4.643200         \\
			0  &  0 & 1/2 & $+--$        & 2.012952 &                1.300898 &       0.253160   &  4.501854         \\
			0  &  1/2 & 0 & $-++$         & 1.999450 &                1.297448 &       0.256070   &  4.643192         \\
			1  &  0 & 0 & $+++$             & 2.616370 &                1.689224 &       0.318154   &  10.681680        \\ \hline
			0  &  0 & 1 & $+++$        & 1.823865 &                1.254068 &       0.306765   &  35.791256        \\
			0  &  0 & 1 &   $+--$    & 1.823419 &                1.253798 &       0.306647   &  35.704179        \\
			0  &  1/2 & 1/2 & $-+-$    & 1.811186 &                1.250473 &       0.310173   &  36.825177        \\
			0  &  1/2 & 1/2 & $--+$    & 1.823421 &                1.253797 &       0.306646   &  35.704160         \\
			0  &  1 & 0 &  $+++$             & 1.963874 &                1.307225 &       0.296636   &  38.197999         \\
			1  &  1/2 & 0 & $-++$         & 2.307335 &                1.483233 &       0.309613   &  58.655102         \\
			1  &  0 & 1/2 & $++-$       & 2.307329 &                1.483234 &       0.309614   &  58.655213         \\
			1  &  0 & 1/2 &   $+-+$1       & 2.322916 &                1.487177 &       0.306095   &  56.869655         \\
			2  &  0 & 0 &  $+++$              & 2.627797 &                1.810987 &       0.398675   &  115.978043         \\ \hline
			0  &  1/2 &  1& $---$   & 1.715167 &                1.226427 &       0.348769   &  170.864714         \\ \hline\hline
	\end{tabular}}
	\end{center}
\end{table}

Another important measure, namely, SRC is the generalization of LMC with definition as follows \cite{src,src2}: $C_{n_r,\ell,m}^{(DD,SRC,\a)}=I_{n_r,\ell,m}^{(DD,2)}\exp\left\{\mathcal{R}_{n_r,\ell,m}^{(DD,\a)}\right\}$. If $\alpha>2$, then $C_{n_r,\ell,m}^{(DD,SRC,\a)}<1$; if $\alpha<2$, then $C_{n_r,\ell,m}^{(DD,SRC,\a)}>1$ and $C_{n_r,\ell,m}^{(DD,SRC,2)}=1$. In the limiting case $\lim\limits_{\a\rightarrow1}C_{n_r,\ell,m}^{(DD,SRC,\a)}=C_{n_r,\ell,m}^{(DD,LMC)}$. This complexity is invariant under scaling transformation \cite{src2}. The results are collected in middle segment of Table~\ref{table4.complexity} for H atom, for same states and parities as that of LMC. It is noticed that, it achieves minimum value for the largest $\ell$, which depends on corresponding parity values. Therefore, in each orbital SRC of order $0<\a<2$ is minimum only for ground state which has no node in radial wave function and corresponding quantum numbers are obtained from $\ell(s_3)=(n-2m(s_1,s_2)-1)/2$. For $\a>2$, SRC attains maximum value for ground state. It is worthwhile noting that, SRC is minimum for largest $\ell$ when $0<\a<2$, and maximum when $\a>2$, for H atom atom without reflection \cite{src2}. The SRC of ground state in each orbital are plotted in panels (C), (D) of Fig.~\ref{fig6.lmc}, with respect to $\ell$ and $m$, having eight parities for order $\a=3/2<1$. The same nature as that in panels (A), (B) is observed here. If $\alpha>2$, then this feature changes. Once again, results on only H atom suffices, as they are independent on $Z$.  

The GRC is an extension of SRC and defined as: \cite{grc,grc2,grc3,grc4,dn.grc,dn.grc2,dn.grc3} $C_{n_r,\ell,m}^{(DD,GRC,\a,\b)}$ $=\ds\exp\left\{\mathcal{R}_{n_r,\ell,m}^{(DD,\a)}-\mathcal{R}_{n_r,\ell,m}^{(DD,\b)}\right\}$. We know that $C_{n_r,\ell,m}^{(DD,GRC,\a,\b)}\lessgtr1$, if $\a\gtrless\b$. Moreover, $C_{n_r,\ell,m}^{(DD,GRC,\a,\b)}=C_{n_r,\ell,m}^{(DD,SRC,\a)}$, when $\a\ne1,\b=2$ and $C_{n_r,\ell,m}^{(DD,GRC,\a,\b)}=C_{n_r,\ell,m}^{(DD,LMC)}$, if $\a\rightarrow 1,\b=2$. We observe that GRC under DS system satisfies several mathematical properties, like inversion symmetry, monotonicity and universal bound, invariance under translations and rescaling transformations, invariance under replication, near continuity, extremal complexity and so on. In absence of reflection, it has been investigated in \cite{grc,grc3}. For infinite order, we obtain $C_{n_r,\ell,m}^{(DD,GRC,\a,\infty)}=\ds\exp\left\{\mathcal{R}_{n_r,\ell,m}^{(DD,\a)}\right\}\|\rho_{n_r,\ell,m}\|$, where $\|\rho_{n_r,\ell,m}\|=\ds \sup_{\mathbf{r}}\rho_{n_r,\ell,m}(\mathbf{r})$, because $\lim\limits_{\beta\rightarrow\infty}\left[\int\left(\rho_{n_r,\ell,m}(\mathbf{r})\right)^{\b}d\chi\right]^{1/\b}=\sup_{\mathbf{r}}\rho_{n_r,\ell,m}(\mathbf{r})$ \cite{l.debnath}. Moreover, in DS system, $C_{n_r,\ell,m}^{(DD,GRC,\a,\infty)}>1$ for all $\alpha>0$ and $C_{n_r,\ell,m}^{(DD,GRC,\infty,\b)}<1$, for all $\beta>0$. These results are found to be satisfied in Schr\"odinger framework without reflection \cite{grc}. It may be mentioned that $C_{n_r,\ell,m}^{(DD,GRC,\a,\b)}\rightarrow0$ for $\alpha>0$ and $\beta\rightarrow0$ \cite{grc}. The GRC results are tabulated in Table~\ref{table4.complexity}, and depicted in panels (E), (F) of Fig.~\ref{fig6.lmc} with respect to $\ell, m$. 

Another complexity measure is RCR, defined between two density functions having atomic numbers $(Z=Z',Z^{''})$, having the following form: \cite{dn.rcr,dn.rcr2} $C_{(n_r',\ell',m'),(n_r^{''},\ell^{''},m^{''})}^{(DD,RCR,\a,\b)}=\exp\left\{\mathcal{R}_{n_r',\ell',m'}^{(DD,\a)}-\mathcal{R}_{n_r^{''},\ell^{''},m^{''}}^{(DD,\b)}\right\}$. In particular, $C_{(n_r,\ell,m),(n_r,\ell,m)}^{(DD,RCR,\a,\b)}=C_{n_r,\ell,m}^{(DD,GRC,\a,\b)}$. Therefore, LMC, SRC, GRC can be obtained from the definition of RCR. Furthermore, we observe that RCR satisfies certain mathematical properties such as, scaling, replication, near continuous and extremal, in the DS system. All these properties without reflection were studied in our previous work \cite{dn.rcr,dn.rcr2,dn.rcr3}. As mentioned earlier in Secs.~\ref{sec3.2shannon} and \ref{sec3.3renyi-entropy}, entropic moments and Shannon entropies of H-like ions with $Z > 1$ are scaled and deviated from H atom ($Z=1$). Alternatively, the amount of scaled factor $[Z^{(2a+1)(1-\a)}]_{Z=2,3,\cdots}$ and deviation amount $[(2a+1)\ln(Z)]_{Z=2,3,\cdots}$ can be obtained easily from a definition of RCR. The composite measure, i.e., statistical complexities are independent of $Z$, but linear information theoretic measures vary with $Z$. The calculated values of RCR for H atom are recorded in Table~\ref{table4.complexity}. The corresponding parameters are provided in table. As in case of LMC, SRC and GRC, RCR values for all $Z$ are same, and thus we produce only results on H atom. These are graphically shown in panels (G), (H) of Fig.~\ref{fig6.lmc}.

\section{Conclusion}\label{sec5.con}
At first, for the Coulomb potential, eigenvalues and eigenfunctions, radial densities and energy degeneracies have been considered from a solution of the DSE, in case of H-isoelectronic atoms
($Z=1-3$). Then we derive expressions for a bunch of linear information measures, namely, standard deviation, Shannon entropy, R\'enyi entropy for arbitrary quantum states. The obtained results are compared and contrasted with corresponding Schr\"dinger system without reflection operator. Next we proceed for four important complexity measures, like LMC, SRC, GRC and RCR for the Dunkl-Coulomb potential. All these results are presented in tabular and graphical form. The information theoretic results are presented here for the first time, and will hopefully inspire future works in this direction. 

\acknowledgments 
AKR is thankful to DST SERB (sanction order CRG/2023/004463) for financial support. AH thanks CSIR, New Delhi (ref. no. 09/0921(16264)/2023-EMR-I) for an Senior Research Fellowship (SRF). 

\small{

}
\end{document}